\documentclass[a4paper,11pt]{article}
\usepackage[ansinew]{inputenc}
\usepackage{amsthm,amssymb,amsmath,amsfonts,amstext}
\usepackage[english]{babel}
\usepackage{array,latexsym}
\usepackage{setspace,graphicx,lscape,longtable,multirow}

\newcommand{\1}{1\!\!1}

\usepackage{commath,mathrsfs,float,bbm,dsfont,soul,subfigure,tikz}
\usepackage[T1]{fontenc}
\usepackage[round]{natbib}
\usepackage[colorlinks=true, linkcolor=blue, citecolor=blue, filecolor=blue, urlcolor=blue, breaklinks=true] {hyperref}
\usepackage[left=2.5cm,right=2.5cm,top=2.5cm,bottom=2.5cm]{geometry}

\newcolumntype{G}{> {\columncolor[gray]{0.8}}c}

\title{New Bias Calibration for Robust Estimation \\ in Small Areas\thanks{The authors are grateful to the ISTAT and University of Pisa for providing access to the data used in this paper. The views
		expressed here are solely those of the authors.}}
\date{}
\author{Setareh Ranjbar\\ 
	Departement of Operations, HEC, University of Lausanne
	\and
	Elvezio Ronchetti  \& Stefan Sperlich  \\
	Geneva School of Economics and Management, University of Geneva}

\begin{document}
\maketitle
\begin{abstract} \noindent
	Using sample surveys as a cost effective tool to provide estimates for characteristics of
	interest at population and sub-populations (area/domain) level has a long tradition in ``small area estimation''. 
	However, the existence of outliers in the sample data can significantly affect the estimation for areas in which they occur, especially where the domain-sample size is small. Based on existing robust estimators for small area estimation we propose two novel approaches for bias calibration. A series of simulations shows that our methods lead to more efficient estimators in comparison with other existing bias-calibration methods. As a real data example we apply our estimators to obtain \textit{Gini} coefficients in labour market areas of the Tuscany region of Italy, where our sources of information are the EU-SILC survey and the Italian census. This analysis shows that the new methods reveal a different picture than existing methods. We extend our ideas to predictions for non-sampled areas.     
\end{abstract}	

{\bf keywords:} small area estimation, robust estimation, bias calibration, non-linear population parameters. 

\section{Introduction}

Small Area Estimation (SAE) has emerged and developed rapidly in recent years in theory and practice. Nowadays, SAE techniques are used in all kinds of official statistics, ranging from business decisions to attribution of health services or allocation of government funds.  This is
partly due to the high demand of statistics by policy makers on the one side, but also to the 
increasing data availability together with recent computational advances on the other side. Using sample surveys is a cost effective tool to provide estimates for characteristics of interest at population and sub-populations' (area/domain) level. This information, coming along with auxiliary data through administrative channels, is used for a better estimation of domain level parameters. Consequently,
when sample sizes in the individual domains are too small (talking of ``small areas'') to obtain reasonable mean square errors by means of direct estimates, then SAE ``borrows strength from other existing sources of information''. For a comprehensive review on this subject we refer to \cite{Roa15} , \cite{ChamClark2012} and \cite{Long2005}. 

Indirect estimators that are based on an explicit linking model are referred to as model-based estimators. Among these we concentrate on mixed effects models (MEM) with area-specific random effects that try to capture the between area variation beyond what is accounted for by the auxiliary covariates, see  \cite{Rao2008R}, \cite{Datta2009}, \cite{Pratesi2016}, \cite{Jiang2006} and \cite{pfeffermann2013}. 

 SAE techniques are intrinsically sensitive to outliers due to the small samples considered. Therefore, robust estimators has been proposed and developed in this field. The two main streams of research on this topic are the robust version of the EBLUP (REBLUP) proposed by \cite{SR09} based on bounded estimating equations for MEM, and the M-quantile (MQ) approach proposed by \cite{Chamb2006}. The latter captures the between area variation through the estimation of area specific quantiles as coefficients; besides being robust against outliers it avoids problems associated with random effect prediction. For more recent developments on the M-quantile methods see \cite{Salvati2012}, \cite{Pratesi2009}, or \cite{Marchetti2017}.

In robust estimation of finite population parameters, \citeauthor{CH86} in his seminal paper in 1986, distinguished between the projective and predictive estimation. The former refers to the classical robust estimation, where the outliers are down weighted or discarded in the estimation procedure. In contrast, the latter accounts for so-called ``representative outliers'', i.e.\ extreme observations in the sample which are likely to occur also among the non-sampled units. Therefore, a calibration is necessary for the bias that is caused by down weighting or disregarding these observations in the estimation process. A general bias calibration approach for the estimation of the finite population Cumulative Distribution Function (CDF) is proposed by \cite{CD1986}, and its robust version is presented in \cite{WR98}. In SAE, \cite{TMC10} introduced a general approach for the bias correction of existing robust estimators, and \cite{Chambers14} discussed different methods to estimate the mean squared error (MSE) for these bias calibrated estimators. 

In this context, we propose two novel approaches to calibrate the bias for the robust estimators of non-linear parameters. In the first one, we derive the non-linear statistical functional (e.g. {\it Gini}) by using  the empirical CDF $\widehat{F}$ on which the calibration is performed. In the second approach, we first linearise the functional by means of an appropriate approximation to afterwards apply a conventional calibration for linear parameter. Namely, we use the von Mises linear approximation by the Influence Function (IF) of the statistic.  While through series of simulation we show that the  former achieves the lowest MSE the latter offers the least absolute bias in comparison to other bias calibration techniques that are presented. We show later that mainly in situations where the errors come from a highly skewed, heavy tailed distributions, one should use an asymmetric calibration to reflect the data generating process. We observe that in such situations the bias calibration leads to a more efficient estimator if the correction is done using the asymmetric Huber function with data-driven tuning parameter.  
   
In  SAE in the absence of closed form formula or good approximation, the MSE are often estimated by some bootstrap techniques; see among others, \cite{Hall2006par}, \cite{Hall2006non}, \cite{Pfeffermann2012}. In our case as well any appropriate bootstrap method introduced in the SAE literature can be used to estimate the MSE of our robust, bias calibrated estimators.  

We have to emphasize that our proposed methods can be used not only for the {\it Gini} but for any other linear or non-linear parameter in small areas. Its use is strongly recommended where the distribution is skewed and/or heavy tailed.  Applications can be found in many fields, from poverty and inequality to environmental or health data. They can also be used to correct for a bias after having imputed missing data.

Section \ref{sec: Gframe} introduces the general framework and the notation. In Section \ref{sec:calib} we propose two approaches to deal with the non-linear population parameters, together with an asymmetric calibration of the estimates. Section \ref{results} compares the performance of these approaches in a series of simulations with existing methods, and Section \ref{sec:tuning} discusses practical issues like the
optimal tuning for the asymmetric calibration.
In Section \ref{sec:App} we apply the EU-SILC survey and the census in Italy, to estimate the \textit{Gini} coefficients at LMA in the Tuscany region. 
Section \ref{sec:conclud} concludes.

\section{General framework and notations} \label{sec: Gframe}

Consider the entire population (so-called super-population) $\mathcal{U}$ of size $N$, that is partitioned in $d$ mutually disjoint sub-populations $\mathcal{U}_{j}$ of size $N_{j}$, corresponding to our small areas $j=1,\cdots d$. For each area $j$ we observe the outcome of interest $ Y_{ij} $ for a sub-sample, $s_{j}$, of individuals $i=1,...,n_j$ but not for the so-called unsampled subset $r_{j}$  of size $N_{j}-n_{j}$. However, we assume that the auxiliary information $ \mathbf{X} $, is available for all units, providing predictive power for the unobserved part of the population. This assumption could be avoided if we focus only on the linear parameters such as mean or the total of the areas but then the methods would be less general.   The $\mathbf{x_{ij}}$ for individual $i$ in area $j$ is a column vector of dimension $p$ that has 1 as its first component.
We are interested in doing inference on $Y_{ij}$ for the area-level. 
When $n_{j}$ is too small for direct estimation, i.e.\ giving large variances, or are not appropriate for other reasons, then model-based small area estimators are used. They apply a model on the super-population, typically to predict the unobserved $Y_{ij}$ for the subsets $r_j$. 
Being interested in the distribution and the corresponding non-linear parameters (see \cite{TMC10}), we do not consider area-level models like those in \cite{Fay79}, \cite{Dick95} or \cite{Pratesi2008}. Instead, we consider unit level models that link the unit outcomes $y_{ij}$ to the unit-specific covariates $\mathbf{x_{ij}}$, see e.g.\ \cite{Battese1988}.  

As a basic setting, assume that the following mixed effect model (MEM) is in place for the sampled as well as for the unsampled units (i.e.\ without sampling selection bias):
\begin{equation} \label{eq-main}
y_{ij}= \mathbf{x_{ij}}^{T}\boldsymbol{\beta}+\mathbf{z}_{ij}^Tu_{j}+\epsilon_{ij}, \qquad \forall j=1,...,d, \qquad \& \qquad \forall i=1,...,n_{j},
\end{equation}
where $\boldsymbol{\beta}$ is the $p$-dimensional vector of fixed effects, and the $u_j$ the random effects of the same dimension as $\mathbf{z}_{ij} \subset \mathbf{x}_{ij}$. 
In our application we concentrate on the commonly used nested error models, where $z_{ij}$ contains only the $1$. Standard assumptions are 
$u_{j}\overset{i.i.d}{\sim} (0,\sigma_{u}) $, and $\epsilon_{ij}\overset{i.i.d}{\sim} (0,\sigma_{e})$ being individual error terms independent from the random effects. In our setting however, to be more realistic  we deviate from this assumption and allow for error terms that may belong to a heavily skewed distribution with potentially heavy tails for which the mean is not necessarily equal to zero. In addition, heteroscedasticity might be present.

Fitting the model to the sample at hand, one obtains estimates of the model parameters which are used to predict the unobserved $Y_{ij}$. By the substitution principle, once the Cumulative Distribution Function (CDF) for each area is estimated, further distribution related quantities (functional statistics) can be derived. \cite{TMC10} pointed out that the CDF estimate is particularly useful in cases where there are extreme values in the small-area sample data, or if the small-area distribution is highly skewed.
The area-specific true CDF for a finite population in area $j$ can be expressed as: 
\begin{align} \label{eq-CDF-main}
F_{j}(t)& =N_{j}^{-1}\left[ \sum_{i\in s_{j}}\1\{ y_{ij}\le t \} +\sum_{k\in r_{j}} \1\{ y_{kj}\le t\} \right]
 =N_{j}^{-1}\left[ \sum_{i\in s_{j}} \1\{  y_{ij}\le t\} +(N_{j}-n_{j})F_{j}^{(2)}(t)\right] . 
\end{align}
Population parameter that can be expressed as a functional of $F_{j}(t)$ can consequently be estimated as a functional of $\widehat{F}_{j}(t)$. 
In a naive setting we may use a plug-in estimator to obtain 
\begin{equation} \label{F2}
\widehat{F}_{j}(t)=N_{j}^{-1}\left[ \sum_{i\in s_{j}}\1\{  y_{ij}\le t\} +(N_{j}-n_{j})\widehat{F}_{j}^{(2)}(t) \right] , \
\widehat{F}_{j}^{(2)}(t)=\frac{1}{N_{j}-n_{j}}\sum_{k\in r_{j}}\1\{  \widehat{y}_{kj}\le t\} .
\end{equation} 
In this case, the estimation of the distribution is achieved by predicting the unobserved units as $\hat{y}_{kj}$. This may be done by using different prediction methods suggested in the literature such as EBLUP, EB, HB, etc. In the presence of outliers, or heavy tailed distribution, one would replace unobserved $Y_{ij}$ rather by robust predictors. For instance, we could use robust mixed linear models to get an estimate of the model parameters and predict the robust version of EBLUP (REBLUP) introduced by \cite{SR09}. Alternatively one may use the M-quantile approach of \cite{Chamb2006} for estimation, and proceed accordingly.  

However, using the expected value of any of these estimators to predict the outcome for non-sampled units $i$ in area $j$ results in a cumulative bias in the estimation of $\widehat{F}_{j}$. Specifically in the presence of heteroskedastic and/or asymmetric error terms, the bias will not cancel in the sum, see \cite{TMC10}. The problem is even more prominent when there exist representative outliers (\cite{CH86}) because these are extreme observations in the sample which are likely to occur also among the non-sampled units. To account for such bias, a calibration step is needed which has also the side-effect of causing some efficiency gain; see \cite{CD1986,WR98,RAO90,Jiongo2013} for the SAE context. 

A basic bias calibration for the CDF was proposed by \cite{CD1986}:
\begin{equation}\label{CDF_CD}
\widehat{F}_{j}^{CD} (t)= N_{j}^{-1}\left[\sum_{i \in s_{j}}\1\{  y_{ij}<t \} +n_{j}^{-1}\sum_{i \in s_{j}} \sum_{k \in r_{j}}\1\{\widehat{y}_{kj}+(y_{ij}-\widehat{y}_{ij})<t \} \right]   .
\end{equation}
In this case the effect of residuals, $y_{ij}-\widehat{y}_{ij}$, is not bounded. \cite{WR98} extended this idea to obtain a bounded version of the prediction of a finite-population CDF
\begin{equation} \label{CDF_WR}
 \widehat{F}_{j}^{WR} (t)= N_{j}^{-1}\left[\sum_{i \in s_{j}} \1\{  y_{ij}<t \} +n_{j}^{-1} \sum_{i \in s_{j}} \sum_{k \in r_{j}}\1\{  \widehat{y}_{kj}^{Rob}+w_{j}\phi_{j} \{(y_{ij}-\widehat{y}_{ij}^{Rob} ) /w_{j}\} <t\} \right]   ,
\end{equation}
where $\widehat{y}_{ij}^{Rob}$ and $\widehat{y}_{kj}^{Rob}$ are the robust prediction of the observed and unobserved outcome, respectively, in area $j$, and $w_{j}$ is a robust estimate of the scale of the residuals in that area. Here, $\phi_{j}$ is a bounded influence function that can change over different areas. \cite{WR98} focused on one finite population; we extend this to several areas. They illustrate that in order to get a more efficient estimate for a finite population CDF, the truncation constant must change at different quantiles of the CDF, with larger constants for more extreme quantiles. Other calibration approaches in the literature are for instance \cite{RAO90} and \cite{Jiongo2013}.  
We first build upon model (\ref{CDF_WR}), propose a skewed calibration that accounts for asymmetry of the error terms, and  extend this idea to correct for the bias in the linearised version of non-linear parameter estimates, namely the {\it Gini} 
index in our application.

\section{Estimation of non-linear parameters for small domains}\label{sec:calib}

Estimators for linear population parameters such as the mean or the total are well studied in the SAE literature. For estimating non-linear statistical functionals, we introduce two approaches of which the first is based on estimating the area specific CDF, and the second on a linear approximation of the parameter. For the latter we use a von Mises approximation with the Influence Function (IF) of the statistical functional. As we are especially concerned about cases where the distribution is highly skewed with a heavy tail and outliers, we start with a robust estimate of the model parameters and  introduce an asymmetric bias calibration method afterwards.  
 
\subsection{Conditional CDF}

Given an estimate of the small areas' CDF, the calculation of linear or non-linear statistical functionals is straightforward. Other advantages of using the CDF estimation are discussed in \cite{TMC10}. 
This approach is extremely beneficial when the small area outcome is highly skewed and/or contains extreme values.  
In this situation we propose a slightly different area-specific CDF estimator than (\ref{F2}). 
In the classical setting, a point prediction is used for the outcome of each unsampled or say, unobserved unit, and the calibration is done using the average effect of the residuals on that point prediction, see (\ref{CDF_CD}) and (\ref{CDF_WR}). We use the empirical predictive distribution of the unobserved outcomes in the estimation of (\ref{F2}) as follows. For each unobserved of the $N_j-n_j$ units in area $j$ we create a vector of length $n_j$ consisting of elements that are the robust predictor $\hat y_{kj}$ plus the vector of residuals obtained from the observed units, cf.\ Step 3 and Step 4 of the procedure below. We do so to preserve the unexplained variation in the sample. This formulation of CDF estimators incorporates the notion of bias calibration proposed by \cite{CD1986,WR98} but using different weights for the observed and predicted units.
The procedure can be summarized by the following steps:
\begin{itemize}
	\item [Step 1] Given a linear MEM, get robust estimates of the model parameter, i.e.\ fixed effects and variance components, as well as robust predictions of the (random) area effects. 
	\item [Step 2] For all observed units $i\in s_{j}$, the residuals are calculated as
	$$ \widehat{\epsilon}_{ij}=y_{ij}-\widehat{y}_{ij}^{Rob}= y_{ij}- \mathbf{x}_{ij}^{T}\widehat{\boldsymbol{\beta}}^{Rob}+\widehat{u}_{j}^{Rob} \qquad \forall i \in s_{j}. $$
	\item [Step 3] Point prediction for the unobserved units $k\in r_j$ are
	$$ \widehat{y}_{kj}^{Rob} =\mathbf{x}_{kj}^{T}\widehat{\boldsymbol{\beta}}^{Rob} +\widehat{u}_{j}^{Rob} \qquad \forall k \in r_{j} .  $$
	\item [Step 4]  For each unobserved unit,instead of only considering one point estimates we would like to use its entire predictive distribution. To do so the predictive distribution (a vector of predictions) is 
	simulated by adding the vector of residuals to the predicted value:
	$$ \widehat{\mathbf{y}}^{p}_{j}= (\textbf{I}_{(N_{j}-n_{j})}\otimes \mathbf{1}_{n_{j}}) \widehat{\mathbf{y}}_{kj}^{Rob}+\mathbf{1}_{(N_{j}-n_{j})} \otimes\widehat{\mathbf{\epsilon}}_{j}, $$
	where $\textbf{I}_{\kappa}$ is the identity matrix of size $\kappa \times \kappa$, $\mathbf{1}_{\kappa}$  is a vector of ones of length $\kappa$, and $\otimes$ is the Kronecker product. Further, $\widehat{\mathbf{y}}_{kj}^{Rob}$ and $ \widehat{\mathbf{\epsilon}}_{j} $ are point predictors and the vector of residuals in area $j$ respectively. Thus, for $N_{j}-n_{j}$ point predictions, the $n_{j}\times (N_{j}-n_{j})$ vector $\widehat{\mathbf{y}}^{p}_{j}$ of simulated outcomes for $r_j$ is created.
\end{itemize}
Pooling this vector with the observed outcomes,  our estimator of the conditional CDF is
$$ \widehat{F}_{j \mid \widehat{u}_{j}}^{BC} (t )= \frac{1}{n_{j}(N_{j}-n_{j}+1)}\left[\sum_{i \in s_{j}}\1\{ y_{ij}<t\}+\sum_{i \in s_{j}} \sum_{k \in r_{j}}\1\{\widehat{y}_{kj}^{Rob}+(y_{ij}-\widehat{y}_{ij}^{Rob})<t\}\right].$$
The notation $j\mid \hat{u}_{j}$ indicates that these estimators are conditioned on the predicted values of the area effects $\hat{u}_{j}$. The denominator in this formula is the sum of $n_{j}+n_{j}(N_{j}-n_{j})=n_{j}(N_{j}-n_{j}+1)$ indicator functions.
The intuition for this calibration follows \cite{CD1986} who used the average effect of the residuals for calibration, c.f.\ (\ref{CDF_CD}). In our method, the whole variation of the residuals is applied to each unobserved subject. 

Since we are looking for a robust estimator, we impose a bounded truncation function on the residual effect as proposed by \cite{WR98}, though adapted to our problem: 
$$ \widehat{F}_{j\mid \widehat{u}_{j}}^{SBC} (t)= \frac{1}{n_{j}(N_{j}-n_{j}+1)}\left[\sum_{i \in s_{j}} \1\{ y_{ij}<t \}
+\sum_{i \in s_{j}} \sum_{k \in r_{j}} \1 \left\{\widehat{y}_{kj}^{Rob}+w_{j}\phi_{j}(\frac{y_{ij}-\widehat{y}_{ij}^{Rob}}{w_{j}})<t\right\}\right],$$
where $\phi_{j}$ is a symmetric Huber-type influence function with weight $w_j$, typically a robust estimate of 
the scale of residuals in area $j$ like the median absolute deviation (MAD). 
Later on we refer to this estimator as REBLUP-SBC, 
a robust EBLUP with Symmetric Bias Calibration.  
Next, we introduce the use of a skewed Huber function.

\subsection{Taking into account the asymmetry of the outcome distribution} \label{sec:calibAlg}

This proposal could be interpreted as an extension of the calibration method of \cite{WR98}. 
We do not, however, calibrate for the estimation of the model parameters (because we are thinking of representative outliers). Furthermore, we argue that in cases where there is extra knowledge available to the researcher, (s)he should exploit this information to better calibrate the estimated CDF, and thereby its statistical functionals, say $T_j$ for area $j$. For instance when analysing income, wealth or expenditure distributions, it is common knowledge that these are strongly skewed with a heavy tail to the right. One can use this information when predicting the distribution of each domain by applying an asymmetric calibration procedure. This requires two truncation constants for the skewed version of the Huber function:
\begin{equation} \label{AHub}
\psi_{c,\gamma}(r)= \left\{\begin{array}{lll}
-c(\frac{2}{\gamma^{2}+1})&if& r \leq -c ,\\
\frac{2}{\gamma^{2}+1} r & if & -c<r<0 ,\\
\frac{2\gamma^{2}}{\gamma^{2}+1} r  & if & 0 \leq r<c ,\\
c(\frac{2\gamma^{2}}{\gamma^{2}+1})  & if & r>c .
\end{array} \right.
\end{equation}  
Here, $c$ defines the width of the truncation window and $\gamma$ the degree of skewness. Like in the classical case of symmetric calibration, one chooses the optimal $c$ and $\gamma$ by minimizing $MSE(T_{j})$. In the presence of   heteroskedasticity, we recommend to consider area-specific sets $(c_{j}, \gamma_{j})$.   

The idea behind $\psi_{c,\gamma}(.)$ is the general presentation of skewed distributions along \cite{Fernandes98}. The tuning parameter $\gamma$ is always positive; while $\gamma=1$ represent the original Huber function, values greater and smaller than 1 provide left and right skewed windows, respectively. From the definition of $\widehat{F}_{j\mid \widehat{u}_{j}}^{SBC}$, $r$ is a standardized residual divided by a robust estimate of its scale. Several choices of the latter are available. We use  
the one of \cite{Rousseeuw1993} which is based on the absolute pairwise differences of the residuals. It is an alternative to more traditional robust estimates but it performs better for skewed distributions. Looking closer at $\psi_{c,\gamma}(.)$ one can see that this is very similar to the skewed Huber function of \cite{Chamb2006} defining the M-quantile method; namely
$$\psi_{c,q}(r)=2\phi_{c}(r) [ q  \1\{ r>0\}+(1-q) \1\{ r\leq0 \} ],$$
where $\phi_{c}(.)$ is the classical Huber influence function, and $q$ the $q$th quantile of the conditional outcome distribution with
$ q=\frac{\gamma^2}{\gamma^2+1}.$

\begin{figure}[htb]
\centering
\includegraphics[width=1\linewidth]{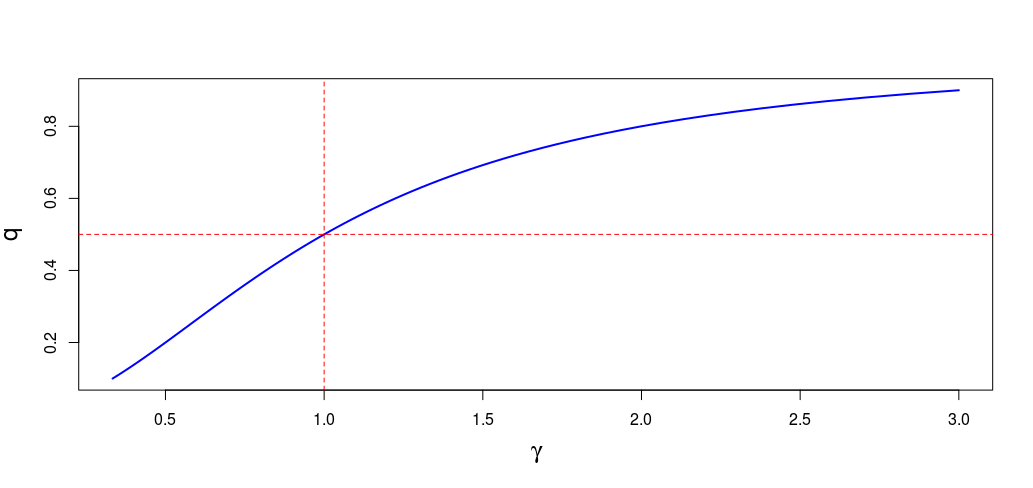}  \label{fig:Gamma-Q}
\caption{The relation between (q) in $\psi_{c,q}(.)$ and ($\gamma$) in $\psi_{c,\gamma}(.)$  }
\end{figure}

Notice however, that here the skewed Huber function is used for calibration, not for estimation. We keep the residuals effect bounded when searching for the shape of the true distribution. In practice, the optimal tuning constants are chosen by considering a mesh of a  $({c,\gamma})$ plane, and estimate the MSE (via bootstrapping) for each combination. As long as one allows for $\gamma=1$, our method also nests symmetric calibration and therefore will outperform it in terms of MSE. 
Provided with the tuning parameter, the area specific CDF estimates are  
\begin{equation} \label{con-dist}
 \widehat{F}_{j\mid \widehat{u}_{j}}^{ABC} (t)= \frac{1}{n_{j}(N_{j}-n_{j}+1)}\left[\sum_{i \in s_{j}}\1\{ y_{ij}<t\} +\sum_{i \in s_{j}} \sum_{k \in r_{j}} \1 \left\{\widehat{y}_{kj}^{Rob}+w_{j}\psi_{c_{j}, \gamma_{j}}(\frac{y_{ij}-\widehat{y}_{ij}^{Rob}}{w_{j}})<t\right\}\right]   
\end{equation}
with $\psi_{c_{j},\gamma_{j}}$ as in (\ref{AHub}) but area-specific.
Functionals like the {\it Gini} index can be calculated subsequently for each area. When the REBLUP is used, we refer to this method as REBLUP-ABC. 

\subsection{Linearization by the Influence Function} \label{sec:IF-BC}

In this section we propose a new alternative way of calibration. It first linearizes the parameter of interest by means of the influence function and then applies the calibration. The idea of using a linear approximations for non-linear parameters has some tradition in SAE, though  mainly for providing an estimator of the variance than correcting for the bias; see \cite{Tille2014,Demnati2004}. 
For the sake of presentation, and because it is the aim of our application, later we
provide an explicit example of the {\it Gini} coefficient.
Consider the first order expansion of a statistical functional introduced by \cite{Mises47}:
\begin{equation} \label{von-Mises}
T(G)-T(F)= \int_{-\infty}^{+\infty} IF(y; T,F)d(G(y)-F(y))+ O\left(\left\lVert G-F \right\lVert^{2}_{2}\right),
\end{equation}
where $F$ is the (model) distribution, $G$ a distribution in its neighbourhood, and $IF(.; T,F)$ the influence function as defined by \cite{Hampel74}. For $G:=\widehat{F}_j$, we get
\begin{equation}\label{eq:linearized}
\widehat{T}_j:= T(\widehat{F}_j)\cong T(F_j)+\frac{1}{N_j}\sum_{i=1}^{N_j} IF(y_{ij};T,F_j)=T(F_j)+\frac{1}{N_j}\sum_{i=1}^{N_j}z_{ij} ,
\end{equation}
where we set $z_{ij}:=IF(y_{ij};T,F_j)$. 
Now, an alternative estimator is obtained by replacing the unknown population parameter in  (\ref{eq:linearized}) with its robust version, 
\begin{equation}\label{eq:linest}
\widehat{T}_j= \widetilde{T}_j+\frac{1}{N_j}\left[\sum_{i \in s_j}z_{ij} + \sum_{k \in r_j} \widehat{z}_{kj} \right] ,
\end{equation}
where $\widetilde{T}_j$ is the original robust estimate of the statistical functional and $\widehat{z}_{kj}:=IF(\widehat{y}_{kj};_jT,F_j)$. 
Substituting robust predictors for all unobserved units, the calibration is done by
\begin{equation} \label{eq:BClin}
\widehat{T}_j= \widetilde{T}_j+\frac{1}{N_j}\left[\sum_{i \in s_j}z_{ij} + \sum_{k \in r_j} \widehat{z}_{kj} + \frac{N_j-n_j}{n_j}\sum_{i \in s_j} w_j \phi(\frac{z_{ij}-\widehat{z}_{ij}}{w_j})\right],
\end{equation}
where $w_j$ is a robust estimate of the scale of the pseudo-residuals 
$\zeta_{ij}=z_{ij}-\widehat{z}_{ij}$ in area $j$, and $\phi(.)$ the Huber function. Hereafter, the result of this calibration approach is referred to as IF-SBC. Noticing that the symmetric calibration is a special case of the asymmetric calibrations we proposed in Section \ref{sec:calibAlg}, we continue with the former for our linearized estimator (\ref{eq:BClin}) by proposing
\begin{equation} \label{eq:IF-ABC}
\widehat{T}_{j}= \widetilde{T}_{j}+\frac{1}{N_{j}}\left[\sum_{i \in s_{j}}z_{ij} + \sum_{k \in r_{j}} \widehat{z}_{kj} + \frac{N_{j}-n_{j}}{n_{j}}\sum_{i \in s_{j}} w_{j}\psi_{c_{j},\gamma_{j}}(\frac{z_{ij}-\widehat{z}_{ij}}{w_{j}})\right]  ,
\end{equation}
with $\psi_{c_{j},\gamma_{j}}$ as in (\ref{AHub}) with area-specific $c_j,\gamma_j$. This bias calibration is referred to as IF-ABC.

\paragraph{Calibration of the Gini coefficient}

As used in our application we compute explicitly the case of \textit{Gini} index.
Among various definitions of this index in the literature, we choose the following that results directly from the classical definition of this index being twice the area between the 45 degrees line and the Lorenz curve:
$$ T(F)= 2 \cdot \frac{I(F)}{\mu(F)}-1$$
where $I=I(F)= \int_{0}^{+\infty}tF(t)dF(t)$, and $ \mu=\mu(F)=\int_{0}^{+\infty}tdF(t).$

Suppressing the area sub-index $j$, the influence function of this functional is:
\begin{equation}\label{eq:IFGini}
IF(y; T,F)= 2 \cdot \big(\frac{1}{\mu}\left[\int_{y}^{+\infty}tdF(t)-I\right]\big) + 2\cdot \frac{y}{\mu}\left[F(y)-\frac{I}{\mu}\right]; 
\end{equation}
see Appendix (\ref{app:IFGini}) for the derivation of (\ref{eq:IFGini}). Now using (\ref{eq:linearized}) in this case we have:

\begin{align*}
\widehat{T} & \cong T(F) +\frac{1}{N}\sum_{i=1}^{N}2 \cdot \big(\frac{1}{\mu}\left[\int_{y_{i}}^{+\infty}tdF(t)-I\right]\big) + 2\cdot \frac{y_{i}}{\mu}\left[F(y_{i})-\frac{I}{\mu}\right]\\
& = T(F) + \frac{-4I}{\mu} + \frac{2}{\mu}\cdot\frac{1}{N}\sum_{i=1}^{N}\left[\int_{y_{i}}^{+\infty}tdF(t)+y_{i}F(y_{i})\right],
\end{align*}

where in the last equality we approximate $\frac{1}{N}\sum_{i=1}^{N}y_{i}$ by $ \mu(F)$ by using the law of large numbers. 
Replacing $T(F)=2\cdot\frac{I}{\mu}-1$, we obtain:

\begin{equation} \label{eq:lin1}
\widehat{T} \cong -T(F)-2 + \frac{2}{\mu}\cdot\frac{1}{N}\sum_{i=1}^{N}\left[\int_{y_{i}}^{+\infty}tdF(t)+y_{i}F(y_{i})\right].
\end{equation}

Letting $z_{i}= \int_{y_{i}}^{+\infty}tdF(t)+y_{i}F(y_{i})$, the non-linear parameter (\textit{Gini} coefficient) in (\ref{eq:lin1}) is  now approximated by a linear function in $z_{i}$, which suggests an alternative estimator for the \textit{Gini} coefficient. Replace the unknown population parameters in equation(\ref{eq:lin1}) with their robust estimates:
$$ \widehat{T}= -\widetilde{T}-2+\frac{2}{\widetilde{\mu}}\cdot\frac{1}{N}\left[\sum_{i \in s}z_{i}+\sum_{k \in r}\widehat{z}_{k}\right].$$
Here the $\widetilde{T}$ and $\widetilde{\mu}$ are the estimates of the \textit{Gini} coefficient and the population mean in which the unobserved units are replaced by robust estimates. 

Using from now on the area-specific notation, and the calibration based on (\ref{eq:IF-ABC}), we have:
\begin{align}\label{IF-ABC}
\widehat{T}^{ABC}_{j}= -\widetilde{T}_{j}-2+\frac{2}{\widetilde{\mu}_{j}}\cdot\frac{1}{N_{j}}\left[\sum_{i \in s_{j}}z_{ij}+\sum_{k \in r_{j}}\widehat{z}_{kj}+ \frac{N_{j}-n_{j}}{n_{j}}\sum_{i \in s_{j}}w_{j}\psi_{c_{j}, \gamma_{j}}(\frac{z_{ij}-\widehat{z}_{ij}}{w_{j}})\right].
\end{align}
Summarizing, the implementation steps for this Gini estimate are:
\begin{itemize}
	\item[] Step 1. Use a robust estimator of the MEM, to get the robust prediction for the unobserved outcomes.
	\item[] Step 2. Use the vector of observed and robustly predicted outcome values in area $j$,  ($\widetilde{Y_{j}}$)  to obtain $\widetilde{T}_{j}$ and $\widetilde{\mu}_{j}$.
	\item[] Step 3. Put $\widetilde{Y_{j}}$ in ascending order and obtain $\widetilde{Y}_{(i)j}$. Then, compute
	$$ z_{ij}= \frac{1}{N}\sum_{h\geq i}\tilde{y}_{(h)j}+\frac{i}{N}\tilde{y}_{(i)j}.$$
	\item[] Step 4. Use (\ref{eq:IF-ABC}) to get the bias calibrated estimates of \textit{Gini} coefficient for each area.
\end{itemize}

\section{Study of Finite Sample Performance} \label{results}

Before applying our new methods REBLUP-ABC and IF-ABC, 
we will validate and compare them in a small simulation study with existing bias calibration methods.
So we aim to estimate the \textit{Gini} coefficients for small areas under different 
designs for model (\ref{eq-main}). Specifically, we generate a population for $ d=40 $ small areas of equal size $N_{j}=300$, $j=1,...,40$ and take a sample of size $ n_{j}=15 $ from each area using SRSWOR (simple random sampling without replacement). The auxiliary variables $X_{ij}$'s are i.i.d.\ with $X_{ij} \sim logNorm(mean=1, sd=0.5)$, and the outcome $Y$ is generated as $ y_{ij}=100+5x_{ij}+ u_{j}+\epsilon_{ij}$, for individual $i$ in area $j$. As we want to study the effect of heavy tailed and right skewed distribution, the error terms are generated from  skewed $t_3$ distributions (St-3) with different measures of skewness to represent the (expected) inequality in developing and developed countries respectively. The scenarios are also created to distinguish between situations where the mean of the heavily skewed error terms is equal to zero, referred to as centred error terms, and those where the mean is different from zero, referred to as  non-centred error terms. Notice that, compared to other measures of inequality, the Gini coefficient is especially sensitive to location changes of the distribution.
However, in all scenarios we keep the distribution of the random area effects as $u \sim N(0,1)$. This is done because \cite{mcculloch2011} argue that in a linear MEM which only contains a random intercept, no association between random effects, error terms and covariates, and uninformative cluster sizes, the misspecification of the shape of the random effects distribution can introduce no or only ignorable bias in the estimation of the model parameters and random components. 
This statement is in accordance with our primary simulations (not shown here). 
The following scenarios are considered here, with $d,N_j,n_j$ as described above, and $u\sim N(0,1)$:  
\begin{enumerate}
	\item Centred errors (we center the $\epsilon$ around zero by subtracting the mean):
	(a) $\epsilon\sim St(3,\lambda= 40)$, (b) $\epsilon\sim St(3,\lambda =70)$, (c) $\epsilon\sim St(3,\lambda =100)$,
	\item  Non-centred error: 
	(a) $\epsilon\sim St(3,\lambda =70)$, (b) $\epsilon\sim St(3,\lambda=150)$, (c) $\epsilon\sim St(3,\lambda =400)$,
\end{enumerate}  
where $St(\cdot, \cdot)$ denotes a right-skewed t distribution, and $\lambda >0$ the measure of skewness as introduced in \cite{Fernandes98}. The symmetric t-distribution is a special case with $\lambda=1$. In these scenarios we choose the $\lambda$ so that the right skewed distributions resemble best the distribution of outcome in different countries, representing the income inequality that is usually observed in developed, developing and underdeveloped regions.
The focus here is mainly to predict area specific parameters, rather than to consistently estimate model parameters. We will see that nevertheless our methods serve as bias calibration techniques that can be applied to any appropriate robust estimation technique.

The data for the population are generated under each scenario and then hundred samples are drawn at random, $t=1 \cdots 100$. We consider the standard (EBLUP), the robust (REBLUP), the REBLUP with symmetric (REBLUP-SBC) and asymmetric (REBLUP-ABC) bias calibration, and two robust bias calibrated method based on the linearized index (IF-SBC, for symmetric calibration) and (IF-ABC, for asymmetric calibration). The \textit{Gini} is predicted and calibrated for each area using each of these techniques. The predicted values are then compared with their true counterparts in relative terms to calculate the relative prediction error:
$$ \text{relative prediction error}(Gini^{(h)}_{j})= \frac{\text{predicted value}(Gini^{(h)}_{j}) - \text{true value}(Gini_{j})}{\text{true value}(Gini_{j})}  . $$
The expected value of these relative errors over repeated sampling provides an estimate of the relative bias in each area. 
$$ \text{Relative Bias}_{j}= \frac{1}{100}\sum_{h=1}^{100}\text{relative prediction error}(Gini^{(h)}_{j}).$$
The prediction of the \textit{Gini} is generally downward biased in small samples. An indirect estimate by replacing the unobserved outcomes with their estimates suffers from the same problem, see \cite{Deltas2003}. That is expected, as the variation in the predicted outcomes (for the unobserved part of the population) is smaller than the variation in true outcomes. Considering the predictive distribution instead of the point prediction as well as skewed truncation corrects somewhat this bias. 

\begin{figure}[htb]
	\begin{center}
		\includegraphics[width=0.9\linewidth,height=20cm]{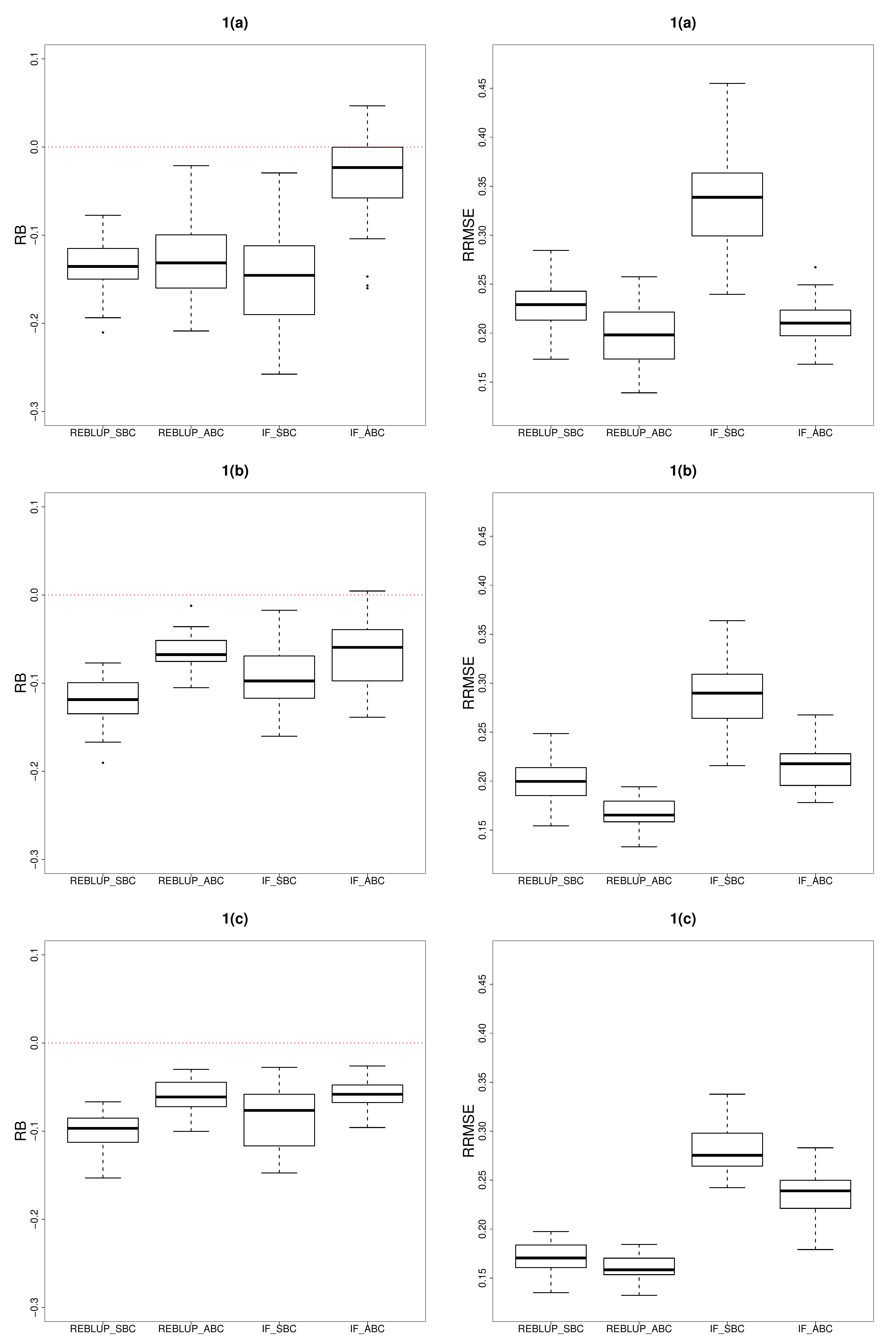}
		\caption{The relative bias and square root relative MSE of the {\it Gini} coefficients, under Scenarios 1(a), (b), and (c)	from top to bottom. }	\label{fig:RRMSE-SC-1}
	\end{center}
\end{figure}

\begin{figure}[htb]
	\begin{center}
		\includegraphics[width=1\linewidth,height=20cm]{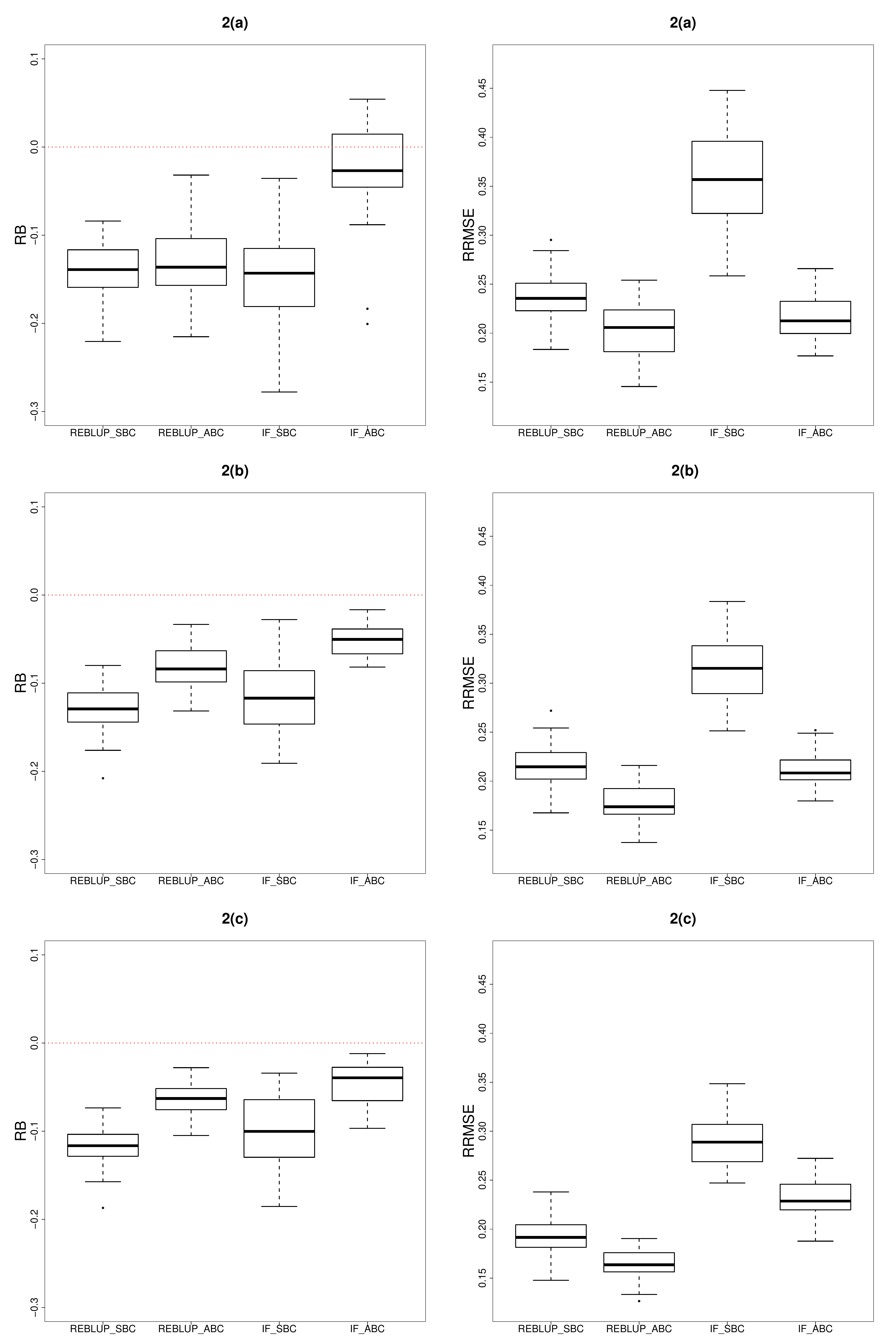}
		\caption{The relative bias and square root relative MSE of the {\it Gini} coefficients, under Scenarios 2(a), (b), and (c) from top to bottom. }	\label{fig:RRMSE-SC-2}
	\end{center}
\end{figure}

To illustrate the efficiency gain due to our proposed methods we compare: 
$$ \text{RRMSE}_{j}=\sqrt{ \frac{1}{100}\sum_{h=1}^{100}\left(\text{relative prediction error}(Gini^{(h)}_{j})\right)^{2}} . $$
Relative Bias and RRMSE for the 40 areas are shown in form of box-plots for each method and scenario, see Figures 
\ref{fig:RRMSE-SC-1} and \ref{fig:RRMSE-SC-2}. Furthermore we summarize the results in Table \ref{RRMSE-Bias} giving the median of Relative Bias and RRMSE over the 40 areas under each scenario and calibration techniques.

	\begin{table}[htb]
		\begin{center}
		\begin{tabular}{c|c|c|c|c|c|c}
			Centred errors
			& \multicolumn{2}{c|}{Scenario (1.a)} & \multicolumn{2}{c|}{Scenario (1.b)} & \multicolumn{2}{c}{Scenario (1.c)}\\ \hline
			median(True Gini)  & \multicolumn{2}{c|}{0.20} & \multicolumn{2}{c|}{0.34} & \multicolumn{2}{c}{0.49}\\ \hline
			Method				& Rel. Bias & RRMSE &  Rel. Bias & RRMSE & Rel. Bias & RRMSE  \\ \hline
			REBLUP				& -0.816 & 0.816 & -0.893 & 0.894 & -0.925 & 0.926 \\  
			REBLUP-SBC & -0.136 & 0.229 & -0.119 & 0.200 & -0.097 & 0.171 \\
			REBLUP-ABC & -0.131 & 0.198 & -0.068 & 0.165 & -0.061 & 0.159 \\
			MQ-SBC &-0.132 & 0.231 & -0.115&  0.204&-0.094 &0.173 \\
			MQ-ABC & -0.110 & 0.220 &-0.087 &0.181 &-0.052 &0.161 \\
			IF-SBC     & -0.146 & 0.339 & -0.097 & 0.290 & -0.076 & 0.275 \\ 
			IF-ABC     & -0.023 & 0.210 & -0.059 & 0.218 & -0.058 & 0.239 \\
			  \multicolumn{7}{c}{  } \\[-2mm]
			Non-Centred errors
			& \multicolumn{2}{c|}{Scenario (2.a)} & \multicolumn{2}{c|}{Scenario (2.b)} & \multicolumn{2}{c}{Scenario (2.c)}\\ \hline
			median(True Gini)  & \multicolumn{2}{c|}{0.20} & \multicolumn{2}{c|}{0.30} & \multicolumn{2}{c}{0.40}\\ \hline
			Method				& Rel. Bias & RRMSE &  Rel. Bias & RRMSE & Rel. Bias & RRMSE  \\ \hline
			REBLUP			   & -0.898     & 0.899     & -0.956     & 0.956     & -0.971     & 0.971   \\ 
			REBLUP-SBC & -0.139 & 0.235 & -0.129 & 0.215 & -0.116 & 0.192 \\ 
			REBLUP-ABC & -0.136 & 0.206 & -0.084 & 0.174 & -0.063 & 0.164 \\ 
			MQ-SBC &-0.135 &0.238 &-0.124 &0.218 &-0.111 & 0.195 \\
			MQ-ABC &-0.109 &0.232 & -0.130 & 0.208 & -0.069 & 0.172 \\
			IF-SBC     & -0.143 & 0.357 & -0.117 & 0.315 & -0.100 & 0.289 \\ 
			IF-ABC     & -0.027 & 0.213 & -0.050 & 0.208 & -0.039 & 0.229  
		\end{tabular}
		\caption{Median of the areas' relative Bias and RRMSE, for REBLUP and different calibration methods (Partial Calibration).}		\label{RRMSE-Bias}
	\end{center}
	\end{table} 

Figures \ref{fig:RRMSE-SC-1} and \ref{fig:RRMSE-SC-2}, and Table \ref{RRMSE-Bias} show that the asymmetric calibration methods are clearly outperforming the symmetric counterparts. IF-ABC provides the best results in terms of Relative Bias, whereas REBLUP-ABC achieves minimum RRMSE for all scenarios. The former finding is not surprising, as the linearization with IF leads to an implicit bias correction. Table \ref{RRMSE-Bias} also gives a summary for calibration based on MQ as an alternative robust method to REBLUP. This is included to have a more comprehensive comparison with other robust estimation methods.

\section{Further Practical Issues} \label{sec:tuning}

Before we apply these methods to our data for estimating the inequality in the different 
LMAs in Tuscany, we need to briefly address two practical issues. The first issue arises when we need to provide a robust prediction for the out of sample areas. In our data set (EU-SILC 2008), out of the 57 LMAs in Tuscany, only 29 are sampled, but 28 are not. The second issue is of technical nature as the proposed methods require two tuning parameters for calibration. In what follows we provide practical solutions for both problems.

\subsection{Full calibration vs.\ partial calibration}\label{Full-BC}

Calibration in our application is done by means of the fitted model residuals. Once a model is assumed to be the Data Generating Process of the super-population, in the absence of sampling selection problems, one can fit it to the sample at hand and predict the unobserved outcomes using the available auxiliary information. The difference between the observed outcomes and the predictions for these observed units is used to calibrate the bias. There are two ways to incorporate these residuals to account for (representative) outliers. One is to use the residuals in each area to correct for the bias in that specific area. This is the framework we considered when we introduced our methods. Let us call it ``Partial Calibration'' because we only use part of the residual vector obtained from the fitted model, namely the area specific residuals.  An alternative is to use each time the entire sample of residuals when calibrating the estimator for one area, say ``Full Calibration''. The latter was also introduced by \cite{Jiongo2013}. However, there are some differences between their calibration, and the way we are using this concept here. In their (full) version, they also try to correct for the bias in the prediction of random effects, whereas in our case, predicted area effects are considered as fixed, because we focus on the conditional CDF of each area. This is also related but still different to what they call ``Conditional Calibration''.
Combining the full calibration idea with our proposed methods which account for the area specific DGP (by choosing area-specific tuning constants for calibration) leads to a compromise
that seems to work well.
We implemented both, partial and full calibration, and compared their results on our application. An advantage of the full calibration technique is that it allows for bias calibration in the non-sampled areas for which we don't have area specific residuals to calibrate in the sense of partial calibration. Now, for (\ref{con-dist}) and (\ref{IF-ABC}) the full calibration analogues are
\begin{align}  \nonumber 
\widehat{F}_{j\mid \widehat{u}_{j}} (t) &= \frac{1}{n_{j}+n(N_{j}-n_{j})}\left[\sum_{i \in s_{j}}I(y_{ij}<t)+\sum_{i \in \bigcup\limits_{h} s_{h}} \sum_{k \in r_{j}}I\left\{\widehat{y}_{kj}^{Rob}+w\psi_{c,\gamma}\left(\frac{y_{ih}-\widehat{y}_{ih}^{Rob}}{w}\right)<t\right\}\right]
\\ \label{full-IFABC}
\widehat{T}_{j} &= -\widetilde{T}_{j}-2+\frac{2}{\widetilde{\mu}_{j}}\cdot\frac{1}{N_{j}}\left[\sum_{i \in s_{j}}z_{ij}+\sum_{k \in r_{j}}\widehat{z}_{kj}+ \frac{N_{j}-n_{j}}{n}\sum_{i \in \bigcup\limits_{h} s_{h}}w\psi_{c, \gamma}(\frac{z_{ih}-\widehat{z}_{ih}}{w})\right]  ,
\end{align}
where  $z_{ij}= \int_{y_{ij}}^{+\infty}tdF_{j}(t)+y_{ij}F_{j}(y_{ij})$, and $w$ a robust estimate of the scale of the entire vector of pseudo-residuals merged together.

\subsection{Choice of the tuning parameters}\label{subsec:tuning}

The choice of tuning parameters can play a crucial role in bias calibration as well as in robust estimation.
Therefore we propose an automatic, data-driven way to find the optimal values for our tuning constants, namely $c$ and $\gamma$, in formula (\ref*{AHub}). In the case of symmetric calibration the convention is to use a rule of thumb for the width of truncating windows. But there also exist some guidelines for the best choice of tuning constants for calibrating certain population parameters, see e.g.\ \cite{WR98}. 

The objective is to minimise the MSE. For estimating the MSE of {\it linear} population parameters some analytic approximations have been proposed, either by first order Taylor expansion,  \cite{PR90}, by defining the estimator as the pseudo linear parameter, \cite{CCT2011}, or by other approximations, \cite{Chambers14}. However, there are mainly two obstacles for our case. First, these approximations do not take into account the calibration, and secondly, there exist no general closed form for non-linear parameters. It is very common then to use resampling methods to approximate the MSE of small area parameter estimates; see e.g.\ \cite{Hall2006par}, \cite{Hall2006non}, \cite{Pfeffermann2012}. We propose to use  non-parametric bootstrap, explained in detail in Appendix \ref{app:boot}. 
This can be used to obtain both, $c$ and $\gamma$. The main drawback of this technique is that it can be computationally expensive. 
When the computational burden becomes too heavy, we suggest to fix $c$ as in the case of symmetric calibration along existing rules of thumb, see \cite{Chambers14}. For the parameter $\gamma$, needed for the asymmetric calibration, we propose as an alternative to the bootstrap, an estimator based on ideas of \cite{Fernandes98}, where
a method of transforming any member of the exponential family into a skewed counterpart is developed. 
This transformation depends essentially on one parameter, which actually corresponds to our $\gamma$ in
(\ref{AHub}), respectively $\gamma_j$ when choosing area-specific tuning constants. 
A simple but effective estimator is
\begin{equation}\label{gammaest} \widehat{\gamma_{j}}=\sqrt{{n_{j}^{-}}/{n_{j}^{+}}}  , \end{equation}
where $n_{j}^{-}$ and $n_{j}^{+}$ are the number of negative and positive centred residuals in area $j$.
When using IF-ABC from equation (\ref{IF-ABC}), these are the residuals of the $z_{ij}$s. 
Appendix \ref{app:gamma} provides some details on the derivation of this formula.

\section{Estimating the {Gini} for Labour Market Areas in the Tuscany}\label{sec:App}

In the following income study, our main interest focuses on the income inequality in LMAs regions of Tuscany, Italy.
We apply the newly developed the robust estimation and bias calibration to estimate the \textit{Gini} index in all LMAs. For this we are provided with the EU-SILC 2008 sample survey of Italy and the 2001 census as an auxiliary source of information.
From the survey we model the household equivalised disposable income on other explanatory variables at household and individual level. Since both EU-SILC sample and census have comparable covariates for individual characteristics, we can exploit the unit level model for our SAE. Specifically, the set of explanatory variables included in this study are gender, marital status, employment status and the years of education of the head of the household (household representative in the survey), as well as household size and household ownership status of the residence.  

LMAs do not match with the administrative boundaries, such that, though graphically and economically of great interest, they are not necessarily a priori considered in the survey planning like for the EU-SILC database. As most of these regions are under-represented in the sample, they must be regarded as small areas. Moreover, from 57 LMAs regions of Tuscany in the census, only for 29 of them we find observations in the sample. That is, for the remaining 28 LMAs direct estimates of the area inequality parameters is not even possible. For these out-of-the-sample areas we use our indirect model predictions calibrating them afterwards by full calibration as explained in Section \ref{Full-BC}.
For the other 29 regions we can alternatively also use partial calibration.
We compare our results obtained when using direct estimators, robust indirect without calibration, using symmetric REBLUP, and when applying our two asymmetric calibration methods, respectively. For the sake of brevity we only show 
some selected results; first when using partial calibration (only for the 29 sampled LMAs available), 
then using full calibration (for comparison reasons applied to all 57 areas, even if only recommended for the 28 
unsampled areas); see Table \ref{tab:popsize} for the number of observations in each area, and its ratio to the population size. In all LMAs less than 1\% of the population is sampled.

The mentioned rules-of-thumb suggest $c=3$ and $c=2$ for REBLUP-ABC and IF-ABC, respectively. 
Figures \ref{fig:App-SBC-sampled} and \ref{fig:App-SBC-full} study the effect of choosing $\gamma$ according to the proposed method (\ref{gammaest}) compared to a range of prefixed alternative values.

\subsection{Results for LMAs in sample areas with partial calibration}

\begin{figure}[htb]
	\begin{center}
		\includegraphics[width=1\linewidth]{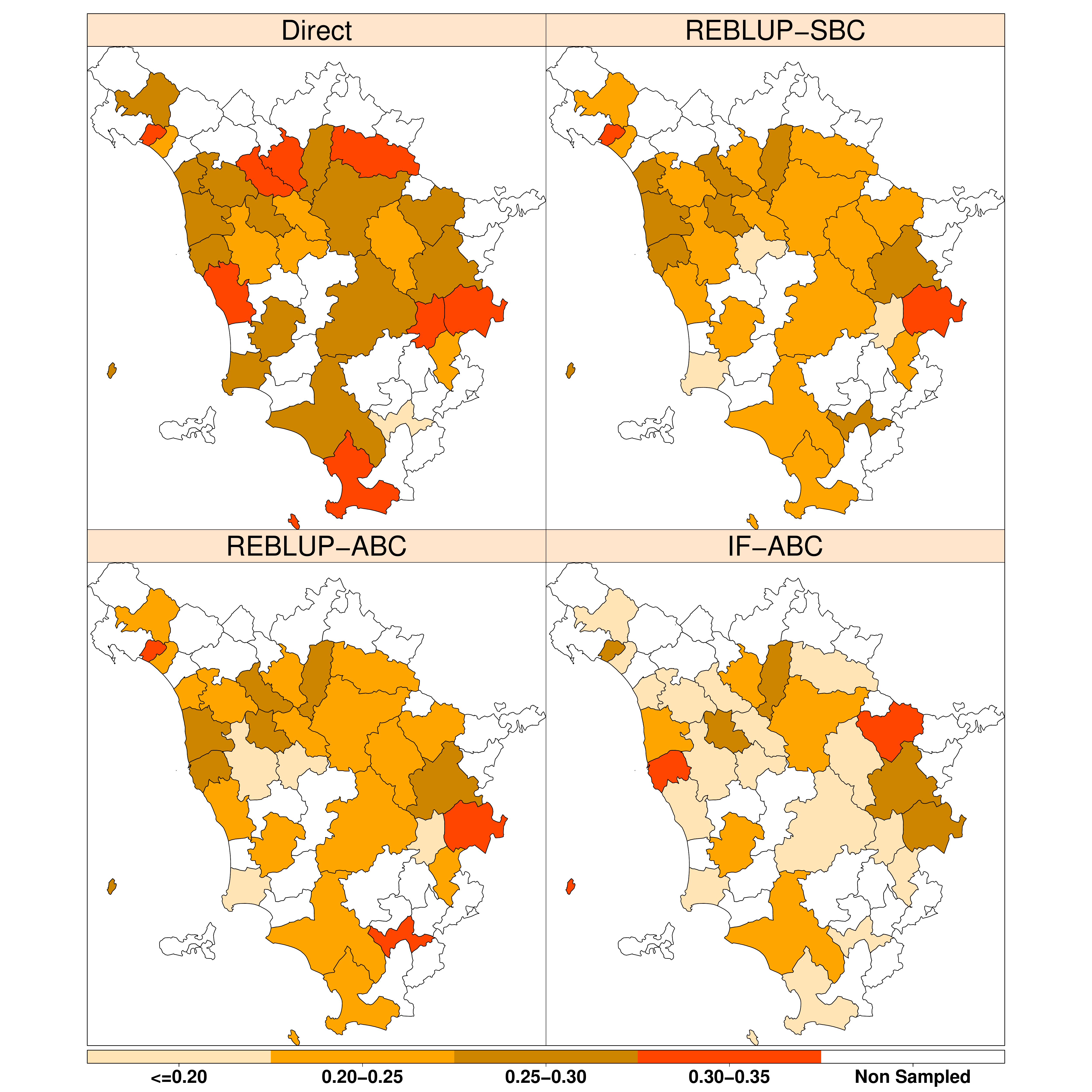}%
		\caption{Gini estimates for the 29 sampled LMAs of Tuscany using different estimation method with partial calibration.}
		\label{fig:App-compare-sampled}
	\end{center}
\end{figure}

\begin{figure}[htb]
	\begin{center}
		\includegraphics[width=.85\linewidth]{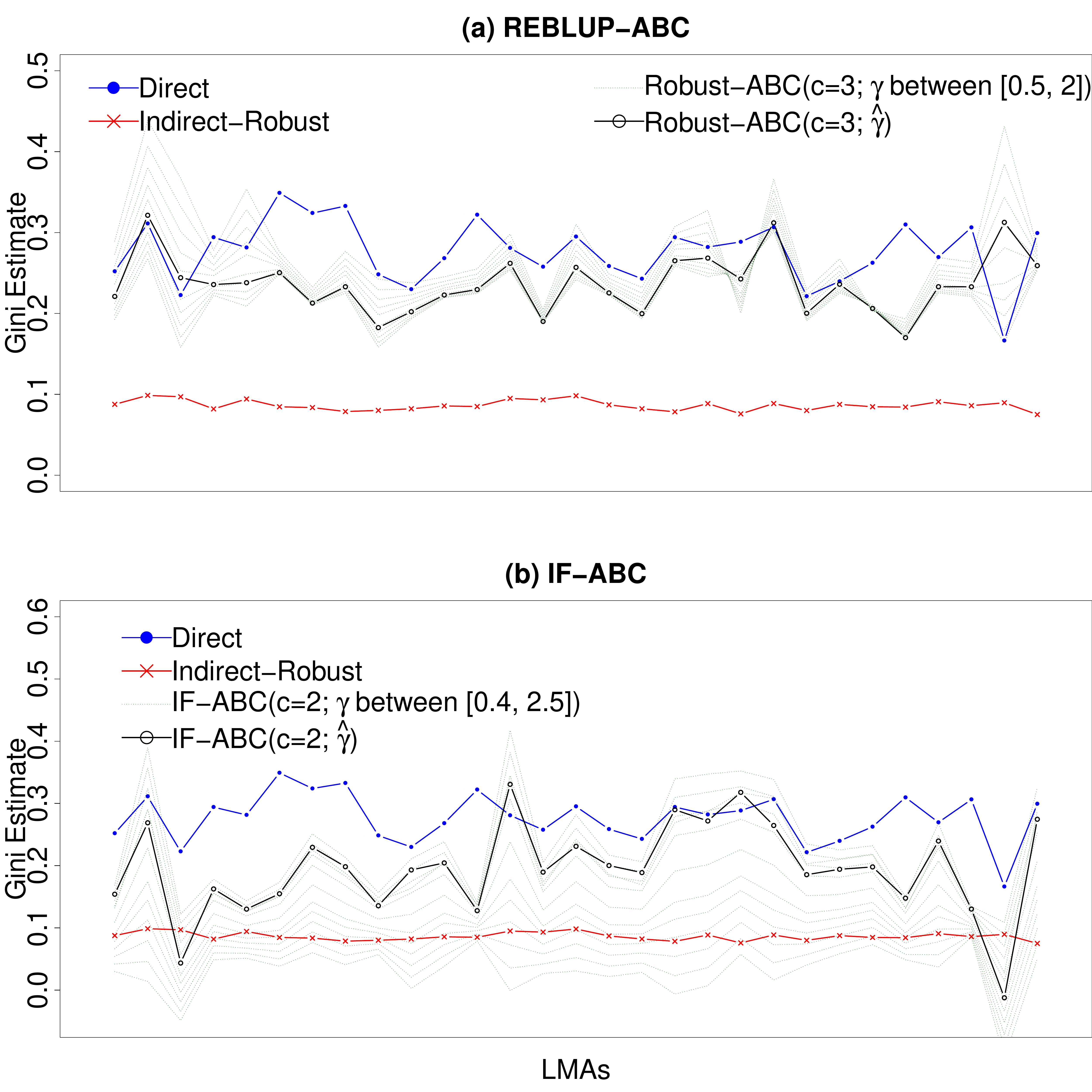}
		\caption{Gini estimates, comparing direct and robust indirect estimators with REBLUP-ABC (upper) and IF-ABC (lower) calibration; the 29 sampled areas using partial calibration.} \label{fig:App-SBC-sampled}  
	\end{center}  
\end{figure}

We first estimate the parameters for the 29 sampled LMAs (recall Table \ref{tab:popsize}) based on the 
presumingly more precise partial calibration. Apart from the robustness study regarding the choice of $\gamma_j$,
Figures \ref{fig:App-compare-sampled} and \ref{fig:App-SBC-sampled} show the differences in the estimation of the \textit{Gini} coefficient due to different calibration methods. Since we do not know the true values of the \textit{Gini} index for each area, we compare the results with the direct estimates which is supposed to be unbiased but with a large variance.
We further compare all this to the indirect estimate that minimizes the variances but inherits some biases. 
While Figure \ref{fig:App-compare-sampled} illustrates on the map, how Gini estimates vary over the different methods, Figure \ref{fig:App-SBC-sampled} shows how asymmetrically bias calibrated estimates change between direct and indirect estimates depending on the choice of tuning parameters. In the former figures we had $c=3$ and $c=2$ for REBLUP-ABC and IF-ABC, and $\gamma_j$ estimated along (\ref{gammaest}). Note that REBLUP-ABC and REBLUP-SBC give quite similar results. It is, however, hard to say whether the latter or IF-ABC are closer to the direct estimates. 
Figure \ref{fig:App-SBC-sampled} shows nicely that our bias calibrated estimators are not just alternatives to direct or robust indirect estimators, but actually offer an extremely useful compromise: 
while still protecting us against the impact of outliers, the bias calibrated estimators maintain a reasonable variance of Gini predictions over areas. The $\gamma$ parameter allows us to smoothly move from one extreme to the other. The estimator (\ref{gammaest}) has a clear
trend towards keeping the bias small, which is typically in the spirit of what practitioners would demand for.

\subsection{Results for all areas with full calibration}

As said, for about half of the LMA we have no observations in the EU-SILC database but only in the census, which in turn does not contain direct information about income. 
When we want to predict now the Gini indices for the 28 unsampled LMAs, one has to switch to the full calibration. This 
can unfortunately imply a heavy smoothing, making all areas looking quite similar unless the distributions of the 
covariates change dramatically over the areas.   
For comparison reasons we give the estimates of the \textit{Gini} coefficients with full calibration for all LMA areas, i.e.\ sampled and non-sampled -- even though in practice one would take partial calibration for the sampled ones. 
In the 28 unsampled areas we predict income for all households by setting the area random effect $\hat{u}_{.}= \hat{u}_{(0.5)}$, which is the median of predicted random effects. Then we use the entire vector of residuals to correct for the bias. For the choice of tuning parameters, $c$ is fixed as before, but $\gamma$ is now estimated once for all areas (i.e.\ not varying over areas) using again the entire vector of residuals for the algorithm introduced in Section \ref{sec:tuning}.

\begin{figure}[htb]
	\begin{center}
		\includegraphics[width=1\linewidth]{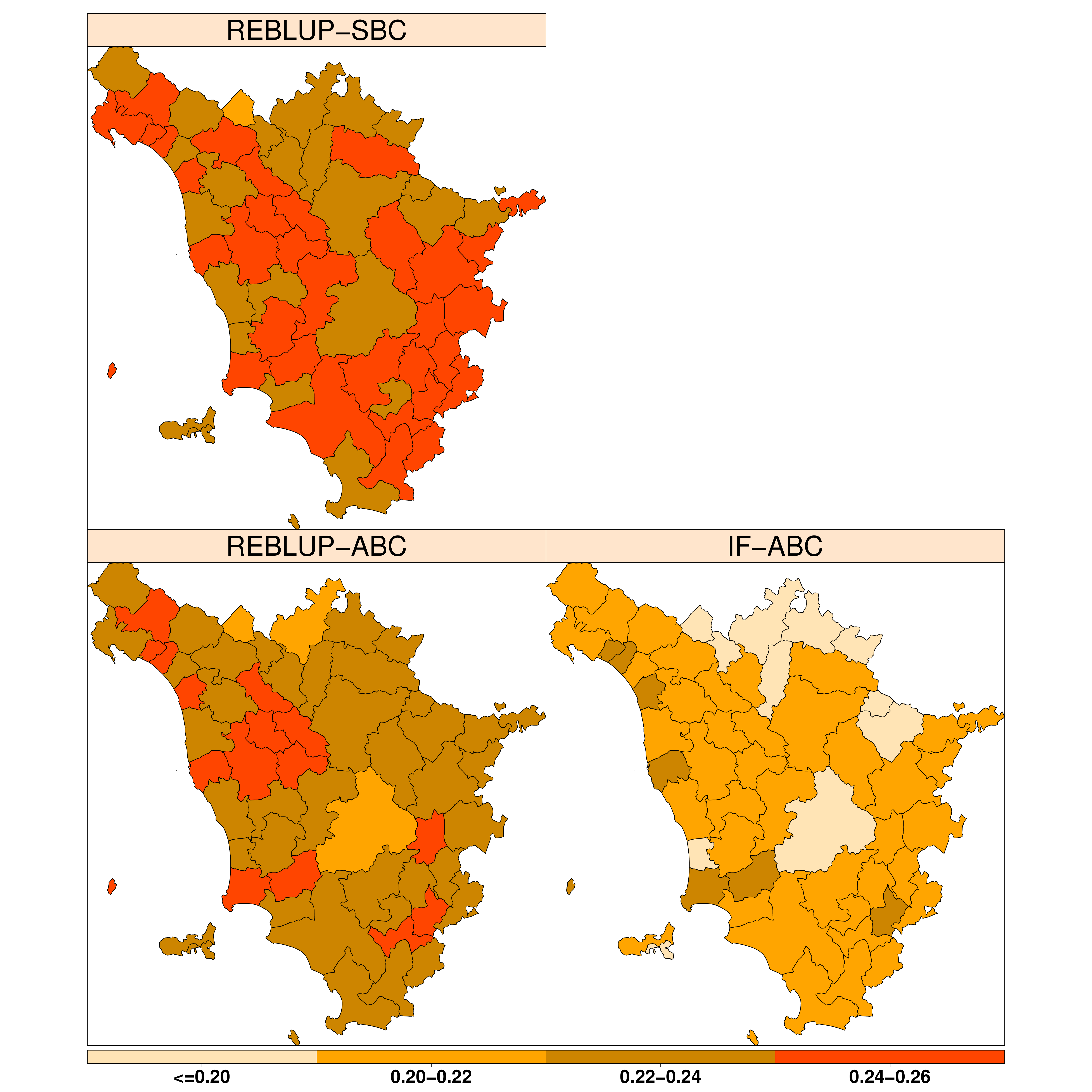}
		\caption{Gini estimates for all 57 LMAs of Tuscany using different estimation methods with full calibration.}
		\label{fig:App-compare-full}  
	\end{center} 
\end{figure}

\begin{figure}[htb]
	\begin{center}
		\includegraphics[width=1\linewidth]{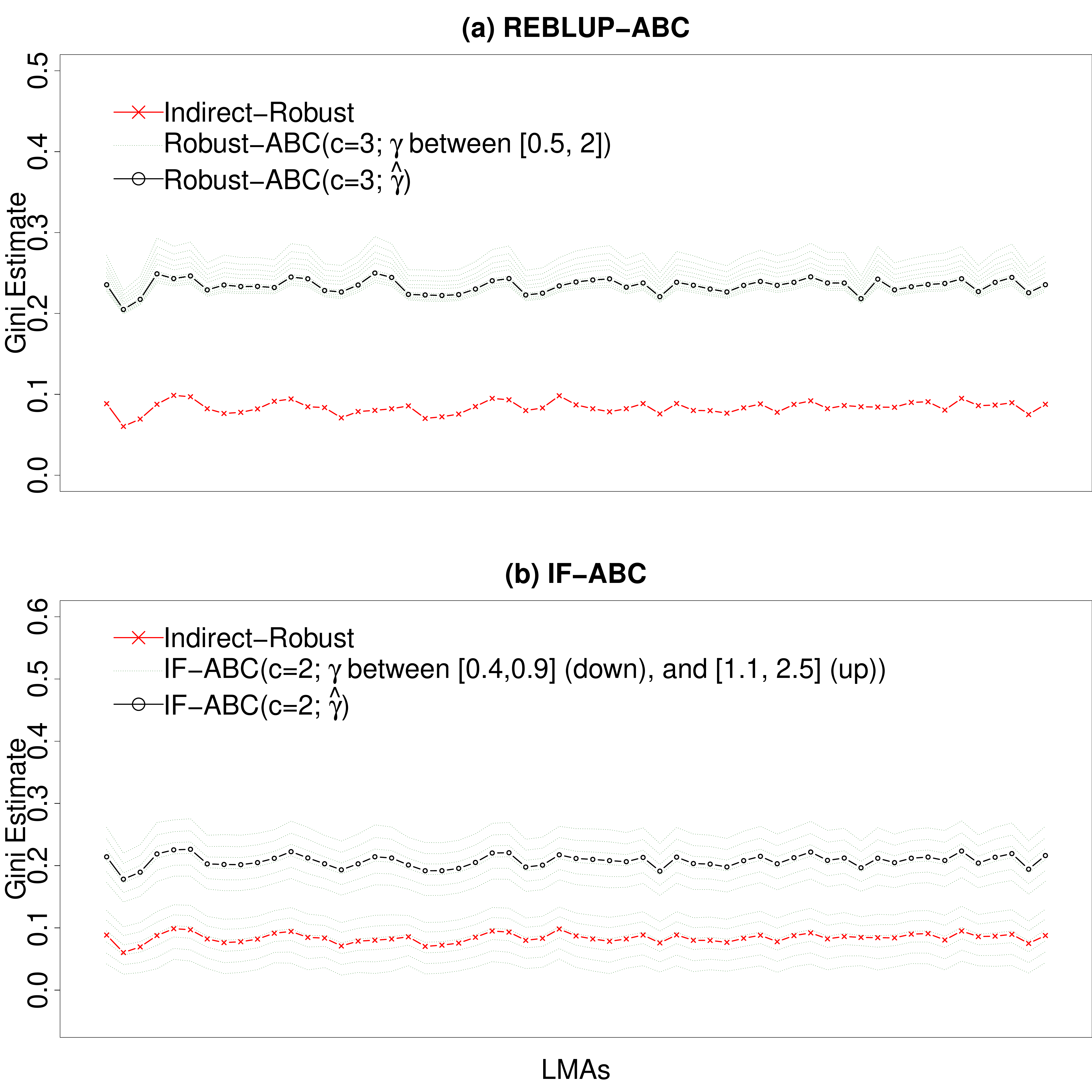}
		\caption{Gini estimates, comparing robust indirect estimators with REBLUP-ABC (upper) and IF-ABC (lower) calibration; all areas using full calibration. }  \label{fig:App-SBC-full}  
	\end{center}  
\end{figure}

In Figures \ref{fig:App-compare-full} and \ref{fig:App-SBC-full}, i.e. the analogues to Figures \ref{fig:App-compare-sampled} and \ref{fig:App-SBC-sampled} (without the direct estimates as these don't exist for unsampled areas), then
we see that the results are strongly smoothed by the full calibration. As said, this is expected since it averages out the calibration over areas by using the entire vector of residuals. Note that the scales in the maps are different from 
those we had for Figure \ref{fig:App-compare-sampled}.
Perhaps surprisingly, here REBLUP-ABC is closer to IF-ABC than to REBLUP-SBC.
Figure \ref{fig:App-SBC-full} is confirming both, what we already found in the context of partial calibration (i.e.\ a strong bias correction effect) as well as what we have seen in Figure \ref{fig:App-SBC-full} (i.e.\ the variation over areas has
been strongly smoothed). Not surprisingly, the effect of the $\gamma$ choice seems to be attenuated, but this may just
be due to the fact that we have taken one value for all areas. 

Taking all our findings together, we can clearly recommend the use of asymmetric bias calibration for the indirect
robust estimators for SAE. One may prefer the calibration via CDF when the aim is to minimize the MSE, but  
the asymmetric calibration through IF if the aim is to minimize the bias. 
In both cases, however, full calibration is only recommended for out of sampled areas to provide some moderate correction of bias in these areas. Because of its strong smoothing effect it should not be used for the sampled areas.
For these, one would take partial calibration and area-wise tuning parameter $\gamma_j$ for which our simple estimator 
(\ref{gammaest}) seems to work fine. Note that this parameter can be used to move from one estreme to the other (between direct estimates and indirect robust ones).

\section{Conclusion} \label{sec:conclud} 

We introduce robust estimators for potentially non-linear small area parameters, and propose various bias calibration approaches to correct for the inherited biases. In the first approach we use an asymmetric Huber function to calibrate the bias of the robust estimator in the CDF in each area, and then apply the bias calibrated CDF to estimate any statistical functional of interest. In the second  approach we derive a linear approximation of the statistical functional using its influence function, and then calibrate for the bias through the linear component. Since the symmetric Huber function is a special case of the asymmetric one, using the latter outperforms symmetric calibration when the tuning parameters are chosen appropriately. Data-driven choices of these parameter are introduced, and modifications of the calibration  allow its application also to those areas in which no sampled units are available. The simulation results confirm the efficiency gain using these approaches compared to the existing methods. While this was mainly shown along the objective of estimating the Gini coefficient, it is clear that these methods can be applied 
to other settings. They provide a way to tackle robust estimation with bias calibration in a fairly general framework. 

After our simulation study which shows the excellent performance of the methods as well as the benefits from applying them, we use these methods to estimate the income inequality for all LMAs in Tuscany, Italy. 
In this application we can clearly see the usefulness of calibration which exhibits quite serious shifts indicating important bias corrections. Also, it shows that full calibration, though extremely useful for doing bias calibration in the non-sampled areas, has a strong smoothing effect. Thus, partial calibration is the preferred choice where doable. 
We also illustrate in a sensitivity check the behaviour of the tuning parameter, and the usefulness of the 
$\gamma_j$ choice(s) via our estimation proposal.

\bibliographystyle{chicago}
\bibliography{BC_bib}

\appendix

\section{Influence function of the \textit{Gini} coefficient}\label{app:IFGini}
Consider the \textit{Gini} coefficient
$T(F)= 2 \cdot \frac{I(F)}{\mu(F)}-1$, where
$ I=I(F)= \int_{0}^{+\infty}tF(t)dF(t)$, $\mu=\mu(F)=\int_{0}^{+\infty}tdF(t)$,  and define
$ F_{\epsilon,y}(t)=(1-\epsilon)F(t)+\epsilon \delta_{y}(t),$
where  $ \delta_{y}(t)= \1 \{t\ge y\} $.
Then 
\begin{align*}
T(F_{\epsilon,y}) &= \frac{2 \int_{0}^{+\infty}t\big((1-\epsilon)F(t)+\epsilon\delta_{y}(t)\big) d\left[ (1-\epsilon)F(t)+\epsilon\delta_{y}(t)\right]}{\int_{0}^{+\infty}t d\left[ (1-\epsilon)F(t)+\epsilon\delta_{y}(t)\right]}-1
\\  & =  \frac{2}{(1-\epsilon)\int_{0}^{+\infty}tdF(t)+\epsilon\int_{0}^{+\infty}t d\delta_{y}(t)} \left\{  (1-\epsilon)^{2}\int_{0}^{+\infty}tF(t)dF(t)  \right. 
\\ & \left. +\epsilon(1-\epsilon)\int_{0}^{+\infty}t\delta_{y}(t)dF(t)
+ \epsilon(1-\epsilon)\int_{0}^{+\infty}tF(t)d\delta_{y}(t)+ \epsilon^2\int_{0}^{+\infty}t\delta_{y}(t)d\delta_{y}(t)
\right\} -1  .
\end{align*}
Using the definition of Dirac delta function we obtain
$$ \int_{0}^{+\infty} tF(t)d\delta_{y}(t)= y \cdot F(y), \ \ \ \ 
\int_{0}^{+\infty} t\delta_{y}(t)d\delta_{y}(t)=y.$$
Therefore, it follows
\begin{equation} \label{eq:G_est}
T(F_{\epsilon,y})= 2\cdot\frac{(1-\epsilon)^{2}I + \epsilon(1-\epsilon)\int_{y}^{+\infty}tdF(t)+\epsilon(1-\epsilon)yF(y)+\epsilon^{2}y}{(1-\epsilon)\mu+\epsilon y}-1,
\end{equation}  
and by definition given in \cite{Hampel74}, the influence function of the functional $T$ is
\begin{align*}
IF(y; T,F) &= \frac{d}{d\epsilon}T(F_{\epsilon,y})\mid_{\epsilon=0} 
= 2 \cdot \frac{\left(-2I+ yF(y) +\int_{y}^{+\infty}tdF(t) \right)\cdot \mu -I\cdot(y-\mu)}{\mu^{2}} \\
&= 2 \cdot \left(\frac{1}{\mu}\left[\int_{y}^{+\infty}tdF(t)-I\right]\right) + 2\cdot \frac{y}{\mu}\left[F(y)-\frac{I}{\mu}\right].
\end{align*}

\section{Nonparametric bootstrap for estimating the RMSE and choosing the tuning constants} \label{app:boot}

Bootstrap procedures are quite popular in the SAE literature as they account for the dependence structure of the data, see for example \cite{Hall2006non} and \cite{Sperlich2010}. Notice, however, that among the nonparametric bootstrap techniques, (x,y)-pair bootstrapping is not feasible due to the clustering of the data.
Given the large literature on bootstrap-based MSE estimation we concentrate here on the conditional RMSE estimation.
We propose a nonparametric bootstrap method that is computationally inexpensive and can help us in
approximating the MSE or RMSE, e.g.\ for determining the tuning constants. 
Since we focus on the conditional distribution for each area (see equation \ref{con-dist}), we will only sample from the error terms and not from the random effects. So once random effects are predicted, we consider them as fixed. This will, indeed, disregard the between area variation and will lead to an underestimation of the RMSE. 
However, bear in mind that the aim is not to provide a precise estimate of the RRMSE per se, but to look for the optimal tuning constants for bias calibration. 

It is well known that the presence of outliers in the residuals can harm the nonparametric residual bootstrapping procedure and lead to misleading inference. This happens when the outliers are overrepresented in the bootstrap sample. To avoid this problem, we do the bootstrap sampling from the pool of winsorised residuals, see \cite{Singh1998} for the breakdown theory of bootstrap quantiles. As our aim is to find optimal constants for the skewed bias calibration function, to evaluate the RRMSE for a given pair $(c_{b}, \gamma_{b})$, we need a, say, 'more relaxed' pair $(c=c_{2}>c_{b},\gamma=1)$, to winsorise the residuals a priori for the bootstrap sampling, see step 4 below. 
More specifically, the bootstrap algorithm is consists then of the following steps. 
\begin{enumerate}
	\item Fit the model with REBLUP; get estimates and predictions for fixed and random parameter. 
	\item Pick a combination of  $c_{boot}$ and $\gamma_{boot}$ from a mesh over a predefined domain of two constants. 
	\item Estimate the bias calibrated \textit{Gini} (or any other inequality index you might be interested in) for each area, say $\widehat{Gini}_{j}^{BC}$, called 'the original estimates' hereafter. Now,
	$\{y_{ij}\}\cup \{\widehat{y}_{kj}\}$ for $j \in s_{j} $ and $ k \in r_{j }$ are considered to be the original population values for area $j$.
	\item Get the residuals for each area from the original fit, and winsorise them as 
	$$ \widehat{res}_{ij}=\psi_{c_{2},1}\left((y_{ij}-\widehat{y}_{ij})/\widehat{w}_{j}\right)\cdot \widehat{w}_{j} , 
	 \quad \mbox{ with } c_{2}> c_{boot},  $$ 
$\widehat{w}_{j}$ are robust estimates of the scale for the residuals of area $j$; e.g. 
 $$ \widehat{w}_{j}= \left(1.4826\times median(\mid \widehat{\boldsymbol{\epsilon}_{ij}}\mid)\right)  .  $$ 
	\item Sample randomly with replacement, separately from each area set of the winsorised residuals, and stack them to build a vector of bootstrap residuals $res_{ij}^{\ast}$.
	\item Construct the bootstrap sample by setting $y_{ij}^{\ast}=\widehat{y}_{ij}+res_{ij}^{\ast}$.
	\item Using this bootstrap sample together with the original design of $x$, fit the bootstrap sample predict the unobserved units, $y^{\ast}_{kj}$, for $k \in r_{j}$.
	\item The bootstrap population outcome set is $\mathcal{U}^{\ast(b)}_{j}=\{y_{ij}\} \cup \{y^{\ast}_{kj}\}.$
	For this population, estimate the parameters of interest, e.g. $Gini_{j}^{\ast(b)}$ where calibration is done by means of the bootstrap model residuals. 
	\item Repeat steps 4.-7.,  B times, each time calculating the error
	 $ \left(\frac{Gini_{j}^{\ast(b)}-\widehat{Gini}_{j}^{BC}}{\widehat{Gini}_{j}^{BC}}\right). $
	\item The estimated RRMSE and Bias are then approximated by
	$$ RRMSE(c_{boot}, \gamma_{boot})= \frac{1}{B}\sum_{b=1}^{B}\left(\frac{Gini_{j}^{\ast(b)}-\widehat{Gini}_{j}^{BC}}{\widehat{Gini}_{j}^{BC}}\right)^{2},$$
	$$ Bias(c_{boot}, \gamma_{boot})=\frac{1}{B}\sum_{b=1}^{B}\left(\frac{Gini_{j}^{\ast(b)}-\widehat{Gini}_{j}^{BC}}{\widehat{Gini}_{j}^{BC}}\right).$$
	\item Repeat steps 2.-10. for all sensible combinations of $\{c_{boot}$ and $\gamma_{boot}\}$, and choose the pair $\{c_{boot}, \gamma_{boot}\}$ that gives the smallest RRMSE.
\end{enumerate}

\section{Details on the estimator for tuning parameter} \label{app:gamma}

Let $f(.)$ be a unimodal symmetric distribution around $0$. \cite{Fernandes98} define a class of asymmetric distributions by
\begin{equation}\label{eq:asdist}
p(\epsilon \mid \gamma) = \frac{2}{\gamma+\frac{1}{\gamma}}\left\{f(\frac{\epsilon}{\gamma}) \1_{(-\infty,0)}(\epsilon)+f(\gamma \epsilon) \1_{\left.[0,\infty\right.)}(\epsilon)\right\}.
\end{equation}  
It follows that
$$\frac{Pr(\epsilon<0\mid \gamma)}{Pr(\epsilon\geq 0 \mid \gamma)}=\frac{\int_{-\infty}^{0}\frac{2}{\gamma+\frac{1}{\gamma}}f(\frac{\epsilon}{\gamma})d\epsilon}{\int_{0}^{\infty}\frac{2}{\gamma+\frac{1}{\gamma}}f(\gamma \epsilon) d\epsilon} , $$
and by change of variables
\begin{equation} \label{eq:gamma}
\frac{Pr(\epsilon<0\mid \gamma)}{Pr(\epsilon\geq 0 \mid \gamma)}=\frac{\int_{-\infty}^{0}f(z)\gamma dz}{\int_{0}^{\infty}f(z)\frac{1}{\gamma} dz}=\gamma^2  ,
\end{equation}
where the last equality holds since $f(.)$ is symmetric around $0$. Assume that our model residuals (used for calibration) follow distribution (\ref{eq:asdist}), and then try to estimate the two probabilities involved in (\ref{eq:gamma}) as follows:
$$ \widehat{Pr}(\epsilon<0\mid \gamma)=\frac{n^{-}}{N},$$
and
$$ \widehat{Pr}(\epsilon\geq 0 \mid \gamma)= \frac{n^{+}}{N},$$
where $n^{-}$, $n^{+}$, and $N$ are the number of positive, negative and total residuals. 
Therefore, a heuristic estimation of $\gamma$, i.e.\ the skewness factor for the residuals that are distributed around $0$, is:  $$ \widehat{\gamma}=\sqrt{{n^{-}}/{n^{+}}} . $$
A feasible algorithm to obtain data-driven tuning constants in (\ref{AHub}) is there the following  
\begin{enumerate}
	\item Centre the block of residuals in each area.
	\item   Fix the constant $c$ at a given value. Values between $2$ and $3$ seem to provide a good performance in practice, see \cite{Chambers14}. 
	\item Count the number of positive and negative centred residuals in each area: $n^{+}_{j}$ and $n^{-}_{j}$ for area $j$ and set $\widehat{\gamma}_{j}=\sqrt{{n^{-}_{j}}/{n^{+}_{j}}}$ . 
\end{enumerate}

\section{Description of EU-SILC data}

\begin{table}[htb]
	\centering
	\caption{Population and sample size for the 29 sampled LMAs areas. }
    \label{tab:popsize}
	\begin{tabular}{cccc}
		\hline
		Area & Population size & Sample size & Percentage sampled \\\hline
		1    & 13265           & 75          & 0.57\%             \\
		2    & 26237           & 27          & 0.10\%             \\
		3    & 29875           & 17          & 0.06\%             \\
		4    & 57848           & 80          & 0.14\%             \\
		5    & 45010           & 33          & 0.07\%             \\
		6    & 43300           & 59          & 0.14\%             \\
		7    & 47363           & 73          & 0.15\%             \\
		8    & 18772           & 25          & 0.13\%             \\
		9    & 15081           & 48          & 0.32\%             \\
		10   & 35350           & 35          & 0.10\%             \\
		11   & 281036          & 261         & 0.09\%             \\
		12   & 28929           & 27          & 0.09\%             \\
		13   & 70240           & 59          & 0.08\%             \\
		14   & 23590           & 25          & 0.11\%             \\
		15   & 71461           & 95          & 0.13\%             \\
		16   & 4619            & 25          & 0.54\%             \\
		17   & 38736           & 24          & 0.06\%             \\
		18   & 33258           & 29          & 0.09\%             \\
		19   & 49371           & 57          & 0.12\%             \\
		20   & 11577           & 27          & 0.23\%             \\
		21   & 12511           & 23          & 0.18\%             \\
		22   & 44078           & 118         & 0.27\%             \\
		23   & 10243           & 22          & 0.21\%             \\
		24   & 42662           & 75          & 0.18\%             \\
		25   & 13087           & 26          & 0.20\%             \\
		26   & 38111           & 35          & 0.09\%             \\
		27   & 14204           & 15          & 0.11\%             \\
		28   & 2829            & 13          & 0.46\%             \\
		29   & 92408           & 132         & 0.14\%            \\\hline
		\end{tabular}
		\end{table}

\end{document}